\documentclass[prl,twocolumn,showpacs,showkeys,amssymb,amsmath,aps]{revtex4}
\usepackage{graphicx}
\usepackage{dcolumn}
\usepackage{bm}
\begin{document}
\title{ Flat Energy histogram version for Interacting Growth Walk }
\author{M. Ponmurugan${}^{\star}$, V. Sridhar${}^{\star}$, 
S. L. Narasimhan${}^{\dagger}$ and K. P. N. Murthy${}^{\star, \ddagger}$} 
\affiliation{ 
 ${}^{\star}$Materials Science Division,
 Indira Gandhi Centre for Atomic Research,\\
 Kalpakkam 603102 Tamilnadu, India}
\affiliation{${}^{\dagger}$Solid State Physics Division,
 Bhabha Atomic Research Centre,\\ 
Mumbai 400085 Maharashtra, India}
\affiliation{${}^{\ddagger}$School of Physics,
University of Hyderabad, 
Central University P.O.,\\
Gachibowli, Hyderabad 500 046, Andhra Pradesh, India }
\date{\today}
\begin{abstract}
Interacting Growth Walks is a recently proposed stochastic  model 
for studying the coil-globule transition of linear polymers.
We propose a  flat energy histogram version for Interacting Growth Walk.
We demonstrate the algorithm on  two dimensional square and triangular lattices 
by calculating the density of energy states  
of Interacting Self Avoiding Walks. 
\end{abstract}
\maketitle

Monte Carlo methods have emerged as a powerful and reliable tool for 
simulating several complex phenomena in statistical physics,
see, {\it e.g.,} \cite{landau,KPN}. Conventional  
Monte Carlo methods \cite{landau} are found  inefficient for
simulating models with complicated energy landscapes, such as 
spin glasses or  protein folding.
There are  many variants of the sampling techniques 
proposed to address these issues, see, {\it e.g.,} parallel
tempering \cite{paral}. Several efforts have  focussed on algorithms 
that aim to produce a nearly uniform distribution in one or 
more of the  macroscopic observables,
such as energy or number of particles, within a predetermined range,
which often fall in the class of 
so-called  flat histogram methods, see below.

Flat histogram techniques permit enhanced flexibility in sampling of energy space in that 
the system is given a greater probability for escaping low-lying energy minima as
compared to traditional Boltzmann sampling.
Multicanonical, transition
matrix, and Wang- Landau algorithms \cite{landau1} are some examples.
These approaches have been used extensively
to study liquid crystals, protein folding, and 
polymer phase behaviour  \cite{braz,binder}. 
This letter focusses on a flat energy histogram method 
to study the coil-globule transition of linear polymers.

Long polymer chains in a good solvent have been studied 
extensively over the  last five decades. Substantial 
progress has been made on  elucidation of  macroscopic 
properties as well as scaling behaviour of an isolated 
polymer chain from  lattice models, see {\it e.g.} 
\cite{vander}. Mostly, 
these have been based on self avoiding walks, as they 
incorporate in their basic definition  excluded volume 
effect.

Self Avoiding Walk (SAW) is a random walk 
in which a walk cannot visit a  site more than once \cite{vander}.
In order to study the thermodynamic properties of SAW,
one needs to assign energy to each conformation. A standard
way of doing this for a lattice SAW is to assign an
energy $\epsilon$ to each  nonbonded nearest neighbor (nbNN)
contact  in the walk. So, a conformation with $m$ such
contacts has energy $E=m \ \epsilon$.
A SAW with energy assigned in this fashion is called an 
interacting self-avoiding walk(ISAW) \cite{vander}. For 
the  study of coil to globule transition of linear
homopolymers, one can set $\epsilon = -1$ without loss of 
generality.

 Many conventional SAW Monte Carlo algorithms \cite{pivot}
and their variants  have been 
formulated for obtaining equilibrium properties of ISAW.
In a recent paper Rechnitzer and  Rensberg \cite{rech} 
define a  random variable called 'atmosphere'. It is the number of 
possible ways a self avoiding walk of 
length $n$ on a lattice can  proceed to create a walk of length $n+1$.  
They showed that from  the statistics of atmosphere one can  estimate 
the connective constant and entropic exponent \cite{rech}.

More recently Prellberg and Krawczyk interpreted the 
atmosphere as proportional to  Rosenbluth and Rosenbluth (RR) weight \cite{RR}
in a stochastic growth algorithm \cite{flatperm}. 
They show that  RR walks in conjuction with the so-called 'dynamic'  
pruning and enrichment \cite{perm} provides a flat histogram 
Monte Carlo algorithm.

In a recent work \cite{pon2}, we have  generalized the notion 
of  atmosphere to other possible growth algorithms; 
we have shown that  the average of  atmosphere taken  over
ensembles of  different  stochastic growth algorithms give an  
estimate of the density of states of ISAW.
In particular we have considered the Interacting Growth Walk (IGW) \cite{igw1,igw2}.
When RR walks are kineticaly interpreted one gets Kinetic Growth Walk (KGW) \cite{kgw}.
In the same way when the PERM-B walks of Grassberger~\cite{Grassberger-PERM-B} 
are kinetically intepreted one gets Interacting Growth Walk \cite{igw1}.

In IGW  the first step from  the origin to  nearest neighbour
site is taken randomly and independently with probability $1/z$ where $z$ is the
coordination number of the lattice. Subsequent steps are taken with 
local IGW probability $p_j$. Let us say that at $j^{th}$
step, $a_j$  unvisited nearest neighbour sites are available for the 
walk to proceed.
The probability for choosing the $\alpha^{th}_j$ unvisited site is 
given by 
\begin{eqnarray}\label{IGWprob}
p_j (\alpha_j)=\frac{exp[\beta_G m_{nbNN}(\alpha_j) ] }
                    {\sum_{\alpha_j =1}^{a_j} exp[\beta_G m_{nbNN}(\alpha_j)]} ,
\end{eqnarray}
where $\beta_G$ is a model growth parameter;
$m_{{\rm nbNN}}(\alpha_j)$ is the number of  nonbonded nearest 
neighbour contacts the walk would 
make if it were to step onto the $\alpha_j^{th}$ nearest neighbour site.
If the number of unvisited nearest neighbour 
sites is zero at any stage, the walk is terminated
(trapping); the entire walk is then discared 
and one starts all over again.
For a given $\beta_G$, the probability of
generating an $N$ step IGW is given by,
\begin{eqnarray}
P_N  =   B(\beta_G) \times  \prod_{j=1}^{N}
                          p_{j},
\end{eqnarray} 
where $B(\beta_G)$ is the normalization constant.
The atmosphere for this $N$ step IGW is given by \cite{pon2},
\begin{eqnarray}
 A_{N} &=&
   \prod_{j=1}^{N} \frac{1}{p_{j}}.
\end{eqnarray}
  
For $\beta_G=0$, every unvisited nearest neighbour site 
is chosen with equal
probability. Therefore IGW reduces to KGW \cite{kgw}
in the  limit $\beta_G \to 0$ \cite{igw2}. 
For large $\beta_G >0$ the walks  
generated by the IGW are extremely compact.
If $\beta_G < 0$, the walk would prefer 
to step onto that  site that leads to lesser  number of contacts.
Hence for large negative $\beta_G$ the walk generated by the IGW
would be  mostly  extended as compared to those with  $\beta_G=0$.

In this letter we propose a flat histogram Interacting Growth Walk.
We show the  atmosphere,  when averaged over the flat histogram IGW ensemble, 
provides a powerful technique  for estimating the density of states (DOS).

In flat histogram IGW, the growth parameter $\beta_G$ is 
taken as a random variable. 
The fluctuations of $\beta_G$ are so adjusted that it ensures uniform accumulation
of $N$-step ISAWs in all the energy bins. In other words if $H(E)$ denotes the energy 
histogram of $N$-step ISAWs, then the random variation of $\beta_G$ from one step
to the next in a walk and from one walk to the other in the ensemble,
leads to a flat $H(E)$.

The energy of an $n$ step ISAW having $m$ contacts is given by $E_n= m \epsilon$. 
Let $m_n^{max}$ denote the maximum number of contacts possible in an $n$ step ISAW.
Then the number of energy levels is $m_n^{max}+1$. These are indexed by 
$k=1,2,...(m_n^{max}+1)$. Thus $E_n$ can take values from
zero to $m_n^{max} \epsilon$  and the correspoding energy histogram 
is represented by $H_n(k)$. We also define an array $\mathcal{G}_n(k)$,
called atmosphere sum, described below.

In  an $N$-step  flat histogram IGW algorithm, 
$H_n(k)$ and $\mathcal{G}_n(k)$ are initially set  zero 
for all $n (1 \le n \le N)$ and $\forall \ k$. 
In  a simulation run, if  a walk  at $n^{th}$ step  has an energy 
$E$ which corresponds to index $k$, then  $H_n(k)$  and 
$\mathcal{G}_n(k)$ incremented as, 
\begin{eqnarray}\label{increment} 	  	 
H_n(k)&=& H_n(k)+1  \nonumber \\
\mathcal{G}_n(k)&=&\mathcal{G}_n(k)+ A_{n},
\end{eqnarray}
where 
\begin{eqnarray}\label{ flatIGWatmos} 	  	 
 A_{n}&=& \prod_{j=1}^{n} \frac{1}{p_{j}}  
\end{eqnarray} 	  
is the  atmosphere \cite{pon2} for $n^{th}$step for that run and
$p_{j}$ is the probability for $j^{th}$ step which 
depends upon the value of $\beta_G$ at that step. 
The implementation of the above algorithm proceeds as follows.

The first step $(n=1)$  of the walk  starts from the origin 
with probability $1/z$, where $z$ is the coordination number of the chosen 
lattice. For $n=1, \ m_1^{max} =0$ and $A_1=z$; the energy histogram
and atmosphere sum  corresponding to the first step is incremented as 
$H_1(1)=H_1(1)+1$ and $\mathcal{G}_1(1)=\mathcal{G}_1(1)+ z$.
All subsequent steps ($j \ge 2$) are  taken with  probability 
$p_j$ given by Eq. ({\ref{IGWprob}}).
The value of $\beta_G$ for each $n(>1)$ is chosen  
depending  upon the difference between  
$H_n(\eta+1)$  and $H_n(\eta)$,
where $\eta = 1,2,...(m_{n-1}^{max}+1)$.  In order to span all energy levels the  
following procedure is adopted in this algorithm.

Let
\begin{eqnarray}
    \Delta &=&  \left \{ H_n(\eta+1) - H_n(\eta) \right \}  \  \xi  \\
 \mathfrak{B}&=&min \left \{ \ \vert \Delta \vert  \ , \ \beta_G^{max} \  \right \}, 
\end{eqnarray}
where $\xi$ is a random number that takes values uniformly between zero to one and 
$\beta_G^{max}$ is the maximum value assigned to avoid 
overflow/underflow problem. The value of $\beta_G$ is chosen  as 
\begin{eqnarray} 	  	 
 \beta_G = \left\{\begin{array}{lll}
    +\mathfrak{B} &  if  &  \Delta < 0    \\
    -\mathfrak{B} &  if  &  \Delta > 0   \\
   \ \mathrm{0} &     if   &   \Delta = 0.  \\
   \end{array} 
   \right.    
\end{eqnarray}
During each growth ste,p  $\beta_G$  takes a value 
between $-\beta_G^{max}$ and  $\beta_G^{max}$.     
The corresponding $H_n$ and $\mathcal{G}_n$ 
are incremented as in Eq.(\ref{increment}).
The walk is continued till it gets trapped or $n$ reaches $N$. 
The above procedure is repeated for several 
number of Monte Carlo attempts M. Estimator for density of states, $g(E)$,
corresponding to the energy levels $k=1,2,..(m_n^{max}+1)$ of ISAW
$\forall \ n \le N$ is then given by 
\begin{eqnarray} 	  	 
  g^{est}_n(k) = \frac {\mathcal{G}_n(k)}{M}.  
\end{eqnarray}

From the density of states we can calculate the 
canonical partition function $Z= \sum_E g(E) \exp(-\beta E)$ 
where $\beta=1/k_BT$. From the partition function 
one can calculate any thermodynamic property.

A general property of IGW is that with increase of $\beta_G$ 
longer walks can be generated with increasing probabilities \cite{igw2}.
In otherwords, attrition is less when $\beta_G$ is large. In fact 
on a two dimensional square lattice, there is no attrition when 
$\beta_G \to \infty$ \cite{igw2,igwhoney}. 
In the context of flat histogram IGW
the parameter that control fluctuation of $\beta_G$ is $\beta_G^{max}$.
Hence if $\beta_G^{max}$  is large, flat histogram IGW can generate longer walks 
spanning uniformly  all the energy levels.

\begin{figure}
\centering
\includegraphics[width=3.0in,height=2.25in]{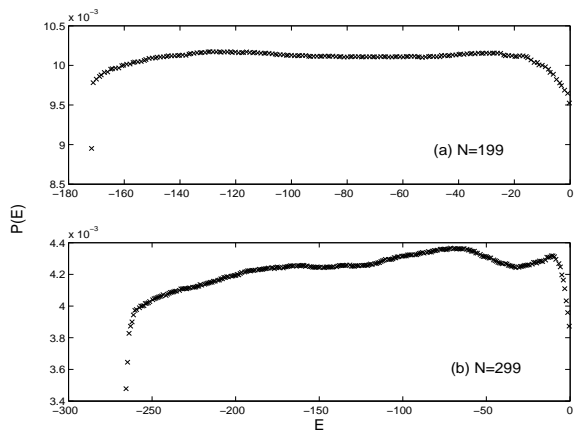}
\caption{Flat histogram IGW with $\beta_G^{max}=10$: Energy distribution for
$N=199$ and $299$. The energy distributions 
are reasonably flat.}\label{flategdis}
\end{figure}

\begin{figure}
\centering
\includegraphics[width=2.5in,height=2.50in]{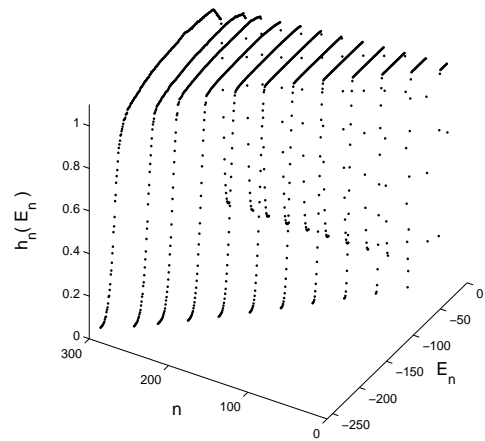}
\caption{Plot of reduced histogram $h_n(E_n)$ versus energy $E_n$ and length 
$n=25$ to $300$ (insteps of $25$: right to left) for ISAW.
These have been  obtained from flat
histogram  IGW on a square lattice with 
$\beta_G^{max}=1.0$ and $M=10^8$ attempts.}\label{flathistsqr}
\end{figure}

\begin{figure}
\centering
\includegraphics[width=2.5in,height=2.5in]{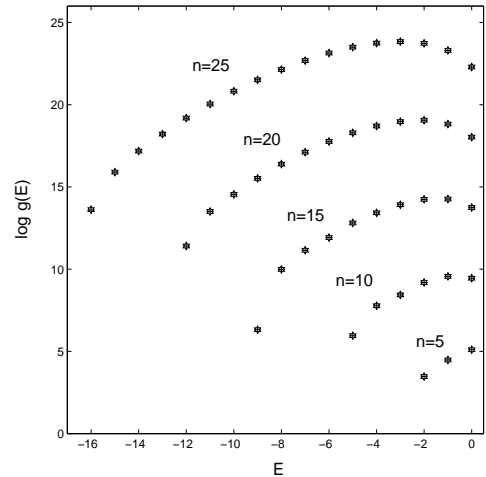}
\caption{Semilog plot of estimated  density of states $g^{est}_n(E_n)$ versus
energy $E_n$ and length $n=5$ to $25$ (insteps of $5$)  
for ISAW on a square lattices obtained from flat histogram IGW with $\beta_G^{max}=1.0$; $M=10^8$.
The Monte Carlo results ($\star$) matches  very well 
with exact (+) density of states. ( Monte Carlo errors are very small; maximum of error
does not exeed one percent and are within the symbol size on the scale).}\label{flatdoscmp}
\end{figure}

\begin{figure}
\centering
\includegraphics[width=2.5 in,height=2.5in]{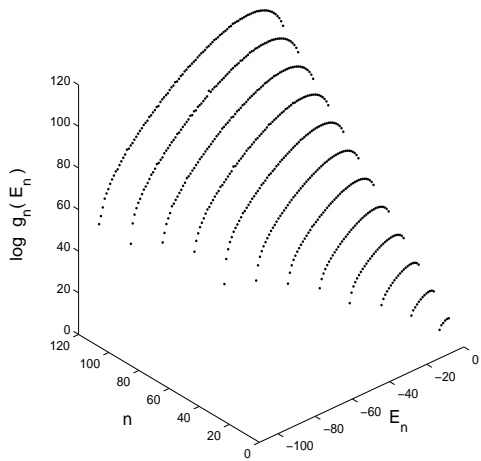}
\caption{Semilog plot of estimated  density of states $g^{est}_n(E_n)$ versus
energy $E_n$ and length $n=10$ to $120$ (insteps of $10$: right to left)
for ISAW on a square lattice obtained from flat histogram IGW with $\beta_G^{max}=1.0$; 
$M=10^8$.}\label{flatdossqr}
\end{figure}

The flat histogram IGW described above was implemented on a two dimensional 
square lattice. Figure \ref{flategdis} depicts the energy 
histogram for $N =199 $ and $299$ for flat histogram IGW  with $\beta_G^{max}=10$.  
We  find that proper choice of $\beta_G^{max}$ would lead better
statistics of density of states.
Thus in our simulation on a square lattice  we have chosen $\beta_G^{max} = 1.0$  and $N \le 300$.   
The reduced histogram $h_n(k)= H_n(k)/ max \{ H_n(k) \} \ \forall k$ for    
$n=25$ to $300$ (insteps of $25$) is shown in Figure \ref{flathistsqr}. 
The histograms are flat.
The estimated density of states for shorter walk length compared with exact enumeration
results is shown in Figure \ref{flatdoscmp}.  The simulation results match  well with 
the exact results. The statistical error of each Monte Carlo data point 
does not exceed $\pm 1\%$. The plot of estimated density of states versus 
energy for walk length $n=10$ to $120$ (insteps of $10$) is shown in 
Figure \ref{flatdossqr}. These results were obtained over $10^8$ 
Monte Carlo attempts.

\begin{figure}
\centering
\includegraphics[width=2.5in,height=2.50in]{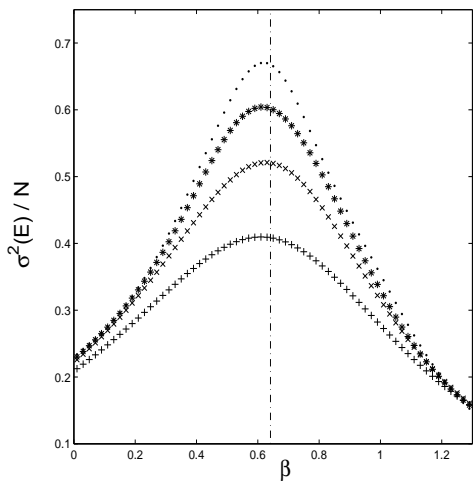}
\caption{Energy fluctuation per walk length  $\sigma^2(E)/N$ versus 
inverse temperature  $\beta$  for ISAW on a square lattice for  
$N=30 (+), 50 (\times), 70 (\ast)$  and  $90 (\cdot)$. 
The vertical lines denote the expected transition temperature 
in the thermodynamic limit $N \to \infty$.}\label{fluct}
\end{figure}

We have calculated fluctuations of energy, 
$\sigma^2(E)=\langle E^2\rangle -\langle E\rangle ^2$  as a function of 
$\beta$. This is shown in Figure \ref{fluct}. 
The  value of $\beta\sim (1/1.54) \sim 0.649$ \cite{vander}
at which phase transition is expected to occur is marked by vertical line. 
The fluctuations are maximum at the transition temperature.

\begin{figure}
\centering
\includegraphics[width=2.5in,height=3.5in]{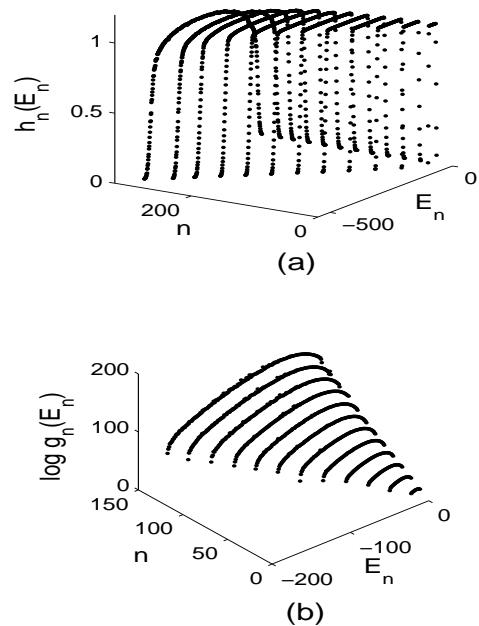}
\caption{Flat histogram IGW on a triangular lattice. (a) Plot of reduced histogram 
(b) Semilog plot of $g_n^{est}(E_n)$  vs $E_n$ for various $n$;
$\beta_G^{max}=0.8$; $M=10^8$.}\label{flattri} 
\end{figure}

The "art" of making flat histogram IGW  to estimate exactly  the DOS of ISAW 
for various  lattices and various dimensions are based to a large extent 
on a suitable choice of $\beta_G^{max}$.
Flat histogram IGW simulation on triangular lattice also were carried out with $\beta_G^{max}=0.8$.
The plot of reduced histogram and density of state  are shown in 
Figure \ref{flattri} . 

In conclusion  we have presented a 
flat energy histogram  method to study Interacting Self Avoiding Walks.
This method is based on IGW algorithm.
Monte Carlo simulation of  flat histogram IGW for longer self avoiding walks 
with large $\beta_G^{max}$  one can obtain resonably flat energy histogram.
By optimizing $\beta_G^{max}$ one can estimate density of states with good statics. 
We have  carried out flat histogram IGW
simulation of  self avoiding walks on square and
triangular lattices
and presents results on DOS of ISAW  and flat energy histogram.

\begin{acknowledgments}
One of the authors (M.P) acknowledges grant from the Council of
Scientific and Industrial Research, India: 
CSIR No : 9/532(19)/2003-EMR-I 
\end{acknowledgments}

\end{document}